\documentclass[aps,pra,twocolumn,superscriptaddress,showpacs,showkeys,amsmath,amssymb]{revtex4}

\usepackage{amsfonts}
\usepackage{amssymb,amsmath}
\usepackage{mathrsfs}
\usepackage{latexsym}
\usepackage{amsmath}
\usepackage[cp1251]{inputenc}
\usepackage{graphicx}
\usepackage{dcolumn}
\usepackage{bm}
\usepackage{color}
\RequirePackage{ifthen}
\RequirePackage[pdfstartview=FitH]{hyperref}

\begin{document}

    \title{$1/N$-expansion for the critical temperature of the Bose gas}
    \author{Orest~Hryhorchak}
    \author{Volodymyr~Pastukhov\footnote{e-mail: volodyapastukhov@gmail.com}}
    \affiliation{Department for Theoretical Physics, Ivan Franko National University of Lviv,\\ 12 Drahomanov Street, Lviv-5, 79005, Ukraine}

    \date{\today}

    \pacs{67.85.-d}

    \keywords{Bose system, critical temperature,
        $1/N$-expansion}

    \begin{abstract}
We revised the large-$N$ expansion for a three-dimensional Bose
system with short-range repulsion in normal phase. Particularly,
for the model potential that is characterised only by the $s$-wave
scattering length $a$ the full numerical calculations of the
critical temperature in the $1/N$-approximation as a function of
the gas parameter $an^{1/3}$ are performed. Additionally to the
well-known result in the dilute limit we estimated analytically
the leading-order strong-coupling behavior of the Bose-Einstein
condensation transition temperature. It is shown that the critical
temperature shift of the non-ideal Bose gas grows at small
$an^{1/3}$, reaches some maximal value and then falls down
becoming negative.
    \end{abstract}

    \maketitle

\section{Introduction}
\label{sec1} \setcounter{equation}{0}

Two decades ago it was clarified \cite{Gruter_Ceperley_Laloe} that
the transition temperature of a three-dimensional homogeneous weakly
interacting system of hard-sphere bosons increases linearly with
the increase of the $s$-wave scattering length
\begin{eqnarray}\label{delta_T_c}
\frac{T_c-T_0}{T_0}=can^{1/3}.
\end{eqnarray}
Soon after its publication this result was confirmed
\cite{Holzmann_Krauth} except for the coefficient $c$. The value
of this coefficient strongly depends on the specific calculation
scheme and was lively debated
\cite{Baym_etal_99,Baym_etal_01,Cruz_Pinto_Ramos,Kneur_Pinto_Ramos,Braaten_Radescu,Wang,Ledowski_Hasselmann_Kopietz,Blaizot,Morawetz_Mannel_Schreiber}.
More precise simulations with classical $\phi^4$ model performed
in Refs.~\cite{Kashurnikov_Prokofev_Svistunov,Arnold_Moore} are
consistent to each other and give for $c\simeq 1.3$. The same
value for the interaction-induced shift of the phase-transition
temperature was also obtained in the later path-integral Monte
Carlo simulations \cite{Nho_Landau}. This discrepancy of different
numerical approaches is discussed in detail in
Ref.~\cite{Pilati_Giorgini_Prokofev}. In the leading order of the
$1/N$-expansion, which effectively sums up particle-hole diagrams,
the coefficient that determines the critical temperature shift is
found to be $c=2.33$ \cite{Baym_Blaizot_Zinn-Justin}. On the other
hand, the summation of the particle-particle ladder diagrams
\cite{Stoof} gives a twice larger value of this coefficient. The
calculations in the so-called fluctuation–exchange approximation
\cite{Tsutsui_Kita} which incorporates both particle-particle and
particle-hole bubbles lead to $c=2.94$. The next-to-leading order
in the $1/N$ term of $c$ lowers the result by $26\%$ (for a single
component Bose system) \cite{Arnold_Tomasik} which particularly
signals a good convergence of the large-$N$ expansion. These
calculations however are complicated by the fact that the first
correction beyond the positive linear shift in $an^{1/3}$ to the
critical temperature is nonanalytic function
\cite{Holzmann_etal,Arnold_Moore_Tomasik,Cruz_etal} and more a
accurate value of $c$ can be obtained only through the five-
\cite{Kleinert}, and seven-loop \cite{Kastening} calculations.
An alternative way to improve the results for the transition
temperature of a dilute Bose gas is to use various resummation
procedures \cite{Yukalovs} where very good agreement with Monte
Carlo simulations is reached.

The goal of the present study is to explore the leading-order
large-$N$ correction to the Bose-Einstein condensation temperature
in the wide region of the interparticle interaction strength and
compare the obtained results with recent Monte Carlo calculations
\cite{Bronin,Nguyen}.

\section{Formulation}
The imaginary-time action of the considered model reads
\begin{eqnarray}\label{S}
S=\int dx\, \psi^*_{\sigma}(x)\left\{\partial_{\tau}+\hbar^2\nabla^2/2m+\mu\right\}\psi_{\sigma}(x)\nonumber\\
-\frac{1}{2N}\int dx\int dx'\Phi(x-x')
|\psi_{\sigma}(x)|^2|\psi_{\sigma'}(x')|^2,
\end{eqnarray}
where $x\equiv(\tau, {\bf r})$, $\int dx=\int_0^{\beta}d\tau\int_Vd{\bf
r}$, and the summation over repeated sort indices $\sigma,
\sigma'=1,2,\ldots ,N$ is understood. As usual \cite{Popov}, the path-integral
is carried out over complex $\beta$-periodic ($1/\beta =T$ is the
temperature) fields $\psi_{\sigma}(x)$. We also introduced the
chemical potential $\mu$ that fixes the total equilibrium density
$Nn$ of the system. The second term of action (\ref{S})
describes interaction between particles with potential
$\Phi(x)=\delta(\tau)\Phi({\bf r})$. For later consideration the
explicit form of the two-body potential $\Phi({\bf r})$ is not
important except it must have the Fourier transform defined.
Introducing new real field $\varphi(x)$ and making use of the
Hubbard-Stratonovich transformation we rewrite action $S$ in the
equivalent form (up to an irrelevant constant term)
\begin{eqnarray}\label{S'}
&&S=\int dx\, \psi^*_{\sigma}(x)\left\{\partial_{\tau}+\hbar^2\nabla^2/2m+\mu-i\varphi(x)\right\}\psi_{\sigma}(x)\nonumber\\
&&-\frac{N}{2}\int dx\int dx'\Phi^{-1}(x-x') \varphi(x)\varphi(x'),
\end{eqnarray}
where the inverse operator $\Phi^{-1}(x-x')$ satisfies natural
identity $\int dx'' \Phi^{-1}(x-x'')\Phi(x''-x')=\delta(x-x')$.
The main advantage of such a decomposition lies in the rearranging
of the perturbation theory in order to stress the leading role of
particle-hole diagrams in the large-$N$ limit. Separating the
uniform classical part of field
$\varphi(x)=\varphi_0+\tilde{\varphi}(x)$ (with constraint $\int
d{\bf r}\tilde{\varphi}(x)=0$) and using the steepest descend method
for the grand canonical potential $\partial \Omega/\partial
\varphi_0=0$ we obtain $i\varphi_0=n\int dx \Phi(x)$, where
density $n$ of each sort of particles should be treated as a
function of chemical potential. Finally, action (\ref{S'}) in
momentum space is
\begin{eqnarray}\label{S_momentum}
S=\sum_{P}\left\{i\omega_p-\varepsilon_p+\tilde{\mu}\right\}\psi^*_{\sigma P}\psi_{\sigma P}+\frac{N}{2}\beta Vn^2\nu_0\nonumber\\
-\frac{N}{2}\sum_{K}\nu^{-1}_k |\varphi_K|^2-\frac{i}{\sqrt{\beta
V}}\sum_{K,P}\varphi_K\psi^*_{\sigma P}\psi_{\sigma P-K},
\end{eqnarray}
where capital letters denote four-momenta $P\equiv(\omega_p, {\bf
p})$,  $K\equiv(\omega_k, {\bf k}\neq 0)$ (here $\omega_k,\omega_p$ are
bosonic Matsubara frequencies) and $\varepsilon_p=\hbar^2p^2/2m$
is free-particle dispersion. We also used notations for the shifted
chemical potential $\tilde{\mu}=\mu-n\nu_0$ and $\nu_k$ for the
Fourier transform of the spherically symmetric inter-particle
potential $\Phi(r)$. Dyson equations that determine both
one-particle Green's function $G_{\sigma}(P)=-\langle\psi_{\sigma
P}\psi^*_{\sigma P}\rangle$ and correlator
$\langle|\varphi_K|^2\rangle$ read
\begin{eqnarray}\label{Dyson_sigma}
G^{-1}_{\sigma}(P)=i\omega_p-\varepsilon_p+\tilde{\mu}-\Sigma_{\sigma}(P),
\end{eqnarray}
\begin{eqnarray}\label{Dyson_varphi}
\langle|\varphi_K|^2\rangle^{-1}=N\left\{\nu^{-1}_k+\Pi(K)\right\},
\end{eqnarray}
where self-energy $\Sigma_{\sigma}(P)$ and polarization operator
$\Pi(K)$ are uniquely determined by vertex function
$\Gamma_{\sigma}(P+K,P)$ (in the following we omit subscript
$\sigma$ everywhere)
\begin{eqnarray}\label{self_en}
\Sigma(P)=\frac{-1}{\beta
V}\sum_{K}\Gamma(P+K,P)\langle|\varphi_K|^2\rangle G(P+K),
\end{eqnarray}
\begin{eqnarray}\label{Pi}
\Pi(K)=\frac{1}{\beta V}\sum_{P}\Gamma(P+K,P)G(P)G(P+K).
\end{eqnarray}
This formulation clearly simplifies the perturbative ($1/N$)
expansion classifying diagrams by the number of
$\varphi$-correlators.

The obtained effective theory constitutes Bose particles coupled
to the field $\varphi_K$ that describes their own collective
excitations. To reveal this connection explicitly it is enough to
calculate the two-point density fluctuations function for each
component of the Bose system
\begin{eqnarray}\label{rho}
\langle|\rho_K|^2\rangle^{-1}=\nu_k+\Pi^{-1}(K),
\end{eqnarray}
which is related to the experimentally measured dynamical
structure factor. Poles of $\langle|\rho_K|^2\rangle$ after
analytical continuation determine the spectrum of collective
modes.

The definition of the shifted chemical potential allows to rewrite
the thermodynamic relation for the equilibrium density as follows
$-\partial \tilde{\Omega}/\partial\tilde{\mu}=Nn$, where
$\tilde{\Omega}=\Omega+\frac{N}{2}Vn^2\nu_0$ depends on temperature
and $\tilde{\mu}$ only. It also naturally enters the
Ward identities in the so-called static limit ($\omega_k=0, {\bf
k}\rightarrow 0$)
\begin{eqnarray}\label{Ward_ident_1}
\frac{\partial G^{-1}(P)}{\partial \tilde{\mu}}=\lim_{K\rightarrow
0}\Gamma(P+K,P),
\end{eqnarray}
\begin{eqnarray}\label{Ward_ident_2}
\lim_{K\rightarrow 0}\Pi(K)=\frac{\partial n}{\partial
\tilde{\mu}}.
\end{eqnarray}
Note that combining the above second identity
(\ref{Ward_ident_2}) and the definition of $\tilde{\mu}$ we find a
full agreement of the low-energy limit of the density-density
correlation function (\ref{rho}) $\lim_{K\rightarrow
0}\langle|\rho_K|^2\rangle=\partial n/\partial\mu$ with the
compressibility sum rule.

\section{Leading-order results}
\subsection{$1/N$-approximation}
We restrict ourselves considering only the simplest approximation of
order $1/N$. In this limit one neglects corrections to the vertex
function, i.e., $\Gamma(P+K,P)=1$ and substitutes bare Green's
function $G^{-1}(P)=i\omega_p-\tilde{\xi}_p$,
($\tilde{\xi}_p=\varepsilon_p-\tilde{\mu}$) in
Eqs.~(\ref{Dyson_sigma}),~(\ref{Dyson_varphi}) for the self-energy
and for the polarization operator. Explicitly writing down the
dependence of the self-energy $\Sigma(P)=\Sigma^{(1)}(P)/N+\ldots$
on $N$ in the leading order we find
\begin{eqnarray}\label{Sigma_(1)}
\Sigma^{(1)}(P)=-\frac{1}{\beta V}\sum_K\frac{\nu_k}{1+\nu_k\Pi(K)}G(P+K),
\end{eqnarray}
where $\Pi(K)=\frac{1}{\beta V}\sum_PG(P)G(P+K)$ is the polarization
operator in the adopted approximation. In order to define the
summation over $\omega_k$ correctly one should carefully single
out the Hartree-Fock (actually, only the Fock) contribution while
calculating $\Sigma^{(1)}(P)$ (see appendix A for details). After
this the derivation of the renormalized one-particle spectrum
becomes a typical routine connected with the analytical
continuation
$\Sigma(P)|_{i\omega_p=\omega+i0}=\Sigma_R(\omega,p)+i\Sigma_I(\omega,p)$
of the self-energy and finding poles of Green's function. Up to
the first order over the expansion parameter $1/N$ the impact of the
interaction effects on the quasiparticle dispersion is given by
$\xi^*_p=\tilde{\xi}_p+\Sigma^{(1)}_R(\tilde{\xi}_p,p)/N$.

We perform the calculation of the Bose-Einstein condensation
transition temperature $T_c$ in a standard way
\begin{eqnarray}
n=-\lim_{\tau \rightarrow +0}\frac{1}{\beta V}\sum_Pe^{i\omega_p\tau}G(P).
\end{eqnarray}
Obviously, in the transition point $\xi^*_p$ tends to zero in the
long-length limit, therefore one has to take into account such a
non-perturbative peculiarity of these calculations. Making use of
spectral representation (via imaginary part $\Sigma_I(\omega,p)$
of the retarded function) for the self-energy and after
straightforward summation over Matsubara frequencies we find in
the accepted approximation
\begin{eqnarray}\label{n}
n=n_0+\frac{1}{N V}\sum_{\bf p}\frac{\partial n(\beta \varepsilon_p)}{\partial \varepsilon_p}\Delta\Sigma^{(1)}_R(\tilde{\xi}_p,p)+\ldots,
\end{eqnarray}
where for convenience we denoted the shifted self-energy
$\Delta\Sigma^{(1)}_R(\tilde{\xi}_p,p)=\Sigma^{(1)}_R(\tilde{\xi}_p,p)-\Sigma^{(1)}_R(-\tilde{\mu},0)$,
introduced notations for the Bose distribution function
$n(y)=(e^y-1)^{-1}$ and for density
$n_0=(p_0/2\sqrt{\pi})^3\zeta(3/2)$ (here and after
$p_0=\sqrt{2mT}/\hbar$, and $\zeta(s)=\sum_{n\geq 1}1/n^s$ is the
Riemann zeta function) of the ideal gas at the critical
temperature. Finally, we conclude this section with the equation
that relates the shift of the critical temperature with the
density jump \cite{Baym_etal_01} in three dimensions
\begin{eqnarray}
\frac{T_c-T_0}{T_0}=-\frac{2}{3}\frac{n-n_0}{n_0},
\end{eqnarray}
where $T_0$ is the Bose condensation temperature of
non-interacting system.

Till now we were discussing the properties of Bose systems not
specifying the form of the inter-particle interaction. The only
requirement to potential was the existence of a well-defined
Fourier transform. In the following sections the calculations of
critical temperature will be performed for a model with a
short-range repulsion, i.e., $\nu_k=4\pi\hbar^2a/m$ which is
characterized only by one parameter, namely the $s$-wave
scattering length $a$.

\subsection{Analytical limits}
We start with the discussion of the dilute limit which is
well-studied due to the long history of investigations. In the
present approach (see appendix A) this limit is reproduced in the
so-called classical approximation, when all Bose distribution
functions in the second term of Eq.~(\ref{n}) are replaced by
their infrared asymptotes $n(y)\rightarrow 1/y$. After these
simplifications, the function that determines the shift of the
critical temperature
\begin{eqnarray}\label{delta_T}
\frac{T_c-T_0}{T_0}=\frac{1}{N}f^{(1)}(an^{1/3})+o(1/N),
\end{eqnarray}
can be calculated analytically in the limit of $an^{1/3}\rightarrow
0$ \cite{Baym_Blaizot_Zinn-Justin}
\begin{eqnarray}\label{f_1_d}
f^{(1)}(an^{1/3})=\frac{8\pi}{3[\zeta(3/2)]^{4/3}}an^{1/3}+\ldots.
\end{eqnarray}
This linear dependence on $an^{1/3}$ appears only because the
large-$N$ expansion even in the simplest approximation sums up
infinite series of diagrams divergent near the phase transition point. 
Any finite order of the conventional perturbation theory
necessarily leads to the incorrect behavior of the critical
temperature shift (for instance, inclusion of the second-order
terms only results in $f^{(1)}(an^{1/3})\propto \sqrt{an^{1/3}}$
\cite{Huang,Bala_Srivastava_Pathak}).

Another highly non-trivial case, where the leading-order behavior
of the function $f^{(1)}(an^{1/3})$ can be obtained analytically,
is the strong-coupling limit $an^{1/3}\gg 1$ \cite{Pastukhov}. To
observe this possibility it is enough to analyze the integral over
$\omega$ in Eq.~(\ref{Sigma_2}). The first multiplier in the
integrand is proportional to the imaginary part of the retarded
density-density correlation function (\ref{rho}). To reveal the
behavior of this correlator we calculated the polarization
operator $\Pi(K)$ (see appendix B for details) and plot the sketch
of the dynamical structure factor
$S(\omega,k)=\frac{1}{\pi}(1-e^{-\beta\omega})^{-1}\Im
\langle|\rho_K|^2\rangle|_{i\omega_k\rightarrow \omega +i0}$ of
this model at fixed coupling constant and for some values of the
wave-vector in Fig.~1.
\begin{figure}
    \centerline{\includegraphics
        [width=0.6\textwidth,clip,angle=-0]{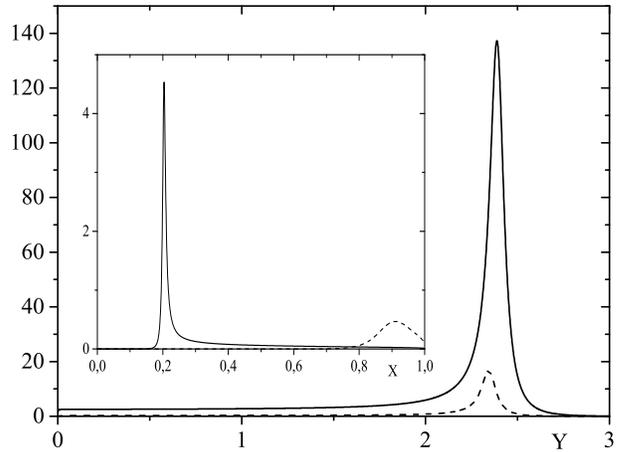}}
   \caption{Dynamical structure factor of a single-component Bose gas in dimensionless units.
   Here the gas parameter $an^{1/3}$ is of order unity and for convenience we introduced dimensionless scaling variables
   \cite{Hofmann_Zwerger} $X=\varepsilon_k/\omega$ and $Y=
   m(\omega-\varepsilon_k)/\hbar^2p_0k$. The solid and dashed lines in main
   panel correspond to $k/p_0=0.01$ and $k/p_0=0.1$, respectively. Inset: $k/p_0=1.0$ (solid) and $k/p_0=10.0$ (dashed).}
\end{figure}
Particularly, it is seen that in the strong-coupling limit the
system possesses a well-defined spectrum of collective modes. It
also easy to show that the damping rate in the phonon region of spectrum decreases exponentially
as the gas parameter increases. Up to the leading order in
$an^{1/3}$ this spectrum of collective excitations can be
approximated by the Bogoliubov one. This excitingly simple fact
allows to provide further calculations analytically and for the
function $f^{(1)}(an^{1/3})$ we obtain (see appendix A) at large
arguments
\begin{eqnarray}
f^{(1)}(an^{1/3})=-\frac{64}{45}\sqrt{\frac{a^3n}{\pi}}+\ldots.
\end{eqnarray}
First note that unlike to the dilute limit (\ref{f_1_d}) the
critical temperature in this case decreases. The second less
noticeable fact is that this leading-order result in the limit
$an^{1/3}\gg 1$ is fully determined by the effects of particle
mass renormalization. On the level of our $1/N$-approximation this
effective mass of a moving particle is nothing but the
hydrodynamic mass of a single impurity atom immersed in the
``phonon'' field $\varphi(x)$. This situation with the function
$f^{(1)}(an^{1/3})$ when it grows at small $an^{1/3}$ and rapidly
falls down to negative values in the strong-coupling limit is
consistent with experiments \cite{Reppy} and with the results of
simulations \cite{Bronin, Nguyen} and has a simple physical
interpretation. For our model the enhancement of the interaction
can be naively treated as an increase of the particle size. In the
ultra-dilute Bose gas where the two-particle collision processes
are less probable the presence of the short-range repulsion only
leads to the reduction of the free volume of the system, that is
why the Bose condensation temperature slightly increases in this
limit. When interaction grows the Feynman mechanism \cite{Feynman}
also takes effect and leads to the effective increasing of a
particle inertial mass. As a consequence, the competition of these
two physical mechanisms of the critical temperature formation
provides that $f^{(1)}(an^{1/3})$ is a non-monotonic function of
$an^{1/3}$.

\subsection{Numerical calculations}
Although formula (\ref{n}) with the self-energy given by
(\ref{Sigma_A}) and (\ref{Sigma_2}) is applicative for the
analytical calculations in diverse limits, the full numerical
computations require another approach. From the practical point of
view it is more convenient to perform primarily integration over
wave-vector in the self-energy and after that to calculate the sum
over the Matsubara frequency. Additionally, the zero-frequency
term should be singled out and treated very carefully
\cite{Baym_Blaizot_Zinn-Justin}. The numerical estimation of the
function $f^{(1)}(an^{1/3})$, which is the main result of the
present paper, is depicted in Fig.~2.
\begin{figure}
    \centerline{\includegraphics
        [width=0.6\textwidth,clip,angle=-0]{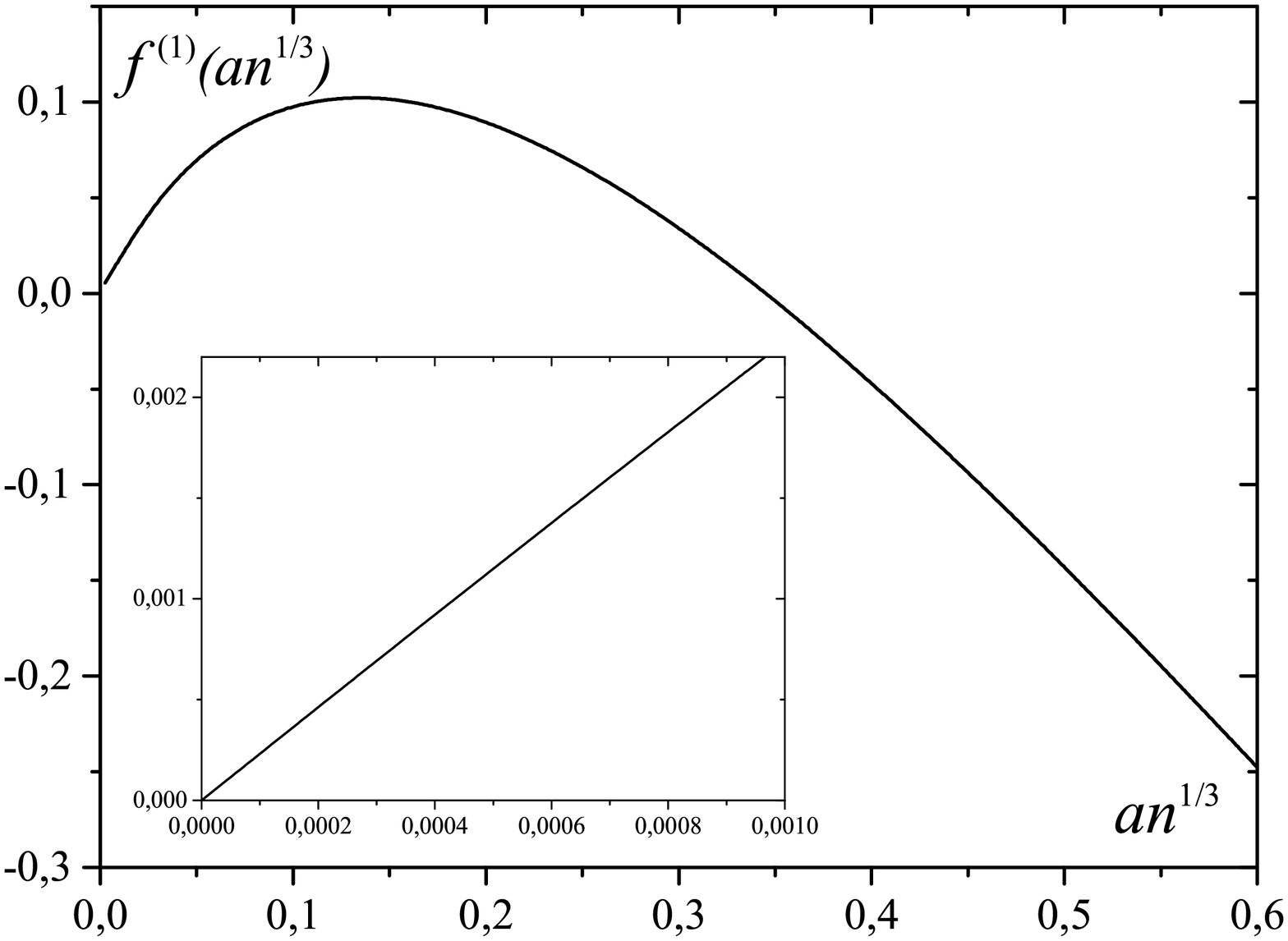}}
   \caption{The result of numerical calculations of the dimensionless $1/N$-shift of
   the critical temperature. The inset shows the linear behavior (\ref{f_1_d}) in the extremely dilute limit.}
\end{figure}
The maximum of the critical temperature due to our
$1/N$-calculations is reached at $an^{1/3}=0.125$ and after
$an^{1/3}=0.345$ the shift of $T_c$ of the Bose system with the
fully repulsive interparticle interaction is negative. For
comparison we also provide in Fig.~3 the results of recent
path-integral \cite{Bronin} and quantum \cite{Nguyen} Monte Carlo
simulations.
\begin{figure}
    \centerline{\includegraphics
        [width=0.6\textwidth,clip,angle=-0]{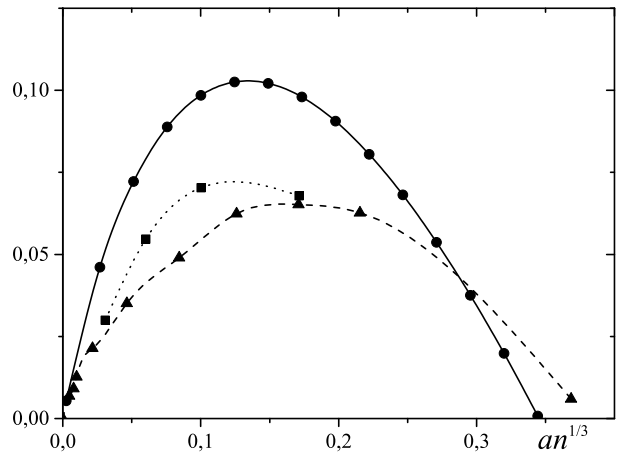}}
   \caption{The comparison of our leading-order $1/N$ calculations (circles) of the critical temperature
   with the results of path-integral \cite{Bronin} (squares) and quantum \cite{Nguyen} (triangles) Monte Carlo simulations.}
\end{figure}

\section{Conclusions}
In summary, by using $1/N$-expansion to the leading order we
performed the full numerical calculations of the critical
temperature for Bose system with short-range repulsion.
Particularly it is shown by the direct computer computations as
well as by the analytical estimations in various limits that the
Bose-Einstein condensation transition temperature of this model is
a non-monotonic function of the gas parameter. We have argued that
for a repulsive inter-particle interaction with the large $s$-wave
scattering length or for a dense Bose gas the critical temperature
shift is negative even in the simplest $1/N$-approximation.
Besides, it is demonstrated that the estimation for the transition
temperature of hard-sphere bosons obtained by means of large-$N$
expansion coincides comparatively well with the essentially exact
results of numerical methods.

\begin{center}
{\bf Acknowledgements}
\end{center}
Insightful discussions with Prof.~Ivan~Vakarchuk and Dr.~Andrij
Rovenchak are gratefully acknowledged. We thank
Prof.~Matthias~Troyer and Dr.~Sebastiano~Pilati for providing us
with the results of their quantum Monte Carlo simulations. This
work was partly supported by Project FF-30F (No.~0116U001539) from
the Ministry of Education and Science of Ukraine.

\section{Appendices}
\subsection{Self-energy}
The summation over the Matsubara frequency $\omega_k$ in
self-energy (\ref{Sigma_(1)}) is not well-defined. To deal with
this problem we single out the leading-order ultraviolet term of
the fraction and rewrite $\Sigma^{(1)}(P)$ in the equivalent form
\begin{eqnarray}\label{Sigma_A}
\Sigma^{(1)}(P)&=&\frac{1}{V}\sum_{{\bf k}\neq 0}\nu_{k} n(\beta \tilde{\xi}_{|{\bf k}+{\bf p}|})\nonumber\\
&+&\frac{1}{\beta V}\sum_K \frac{\nu^2_k \Pi(K)}{1+\nu_k\Pi(K)}G(P+K),
\end{eqnarray}
where the first term is the Fock correction. Now the second sum is
convergent and by using of the spectral theorems can be written in
terms of real and imaginary part of the retarded polarization
operator $\Pi(K)|_{i\omega_k\rightarrow
\omega+i0}=\Pi(\omega,k)=R(\omega,k)+iI(\omega,k)$
\begin{eqnarray}\label{Sigma_2}
\frac{1}{V}\sum_{{\bf k} \neq 0}\int^{\infty}_{-\infty}\frac{d \omega}{\pi}
\frac{\nu^2_k I(\omega,k)}{\left |1+\nu_k\Pi(\omega,k)\right |^{2}}\nonumber\\
\times\frac{n(\beta\omega)-n(\beta\tilde{\xi}_{|{\bf k}+{\bf p}|})}{\omega - \tilde{\xi}_{|{\bf k}+{\bf p}|}+i\omega_p}.
\end{eqnarray}
In the strong-coupling limit, where the dynamical structure factor
of the system has a sharp peak which corresponds to the
quasiparticles with the Bogoliubov-like dispersion
$E^2_k=\varepsilon^2_k+2n\nu_k\varepsilon_k$, one can integrate
over $\omega$ in Eq.~(\ref{Sigma_2}) and obtain in the leading
order:
\begin{eqnarray}
&&\Delta\Sigma^{(1)}_R(\tilde{\xi}_p,p)=\nonumber\\
&&-\frac{4\varepsilon_p}{3V}\sum_{{\bf k}\neq 0}
\frac{n\nu^2_k\varepsilon^2_k}{E_k(E_k+\varepsilon_k)^3}+o(\sqrt{a^3n}).
\end{eqnarray}

\subsection{Polarization operator}
In order to reveal the structure of collective modes of the system
and to calculate the self-energy we give some details of the
polarization operator calculations. The imaginary part of the
retarded polarization operator calculated at $T_c$ reads in the
adopted approximation
\begin{eqnarray}\label{I_omega}
I(\omega, k)=\frac{\beta p^4_0}{16 \pi k}\ln\left
|\frac{1-e^{-[k/2p_0+\beta\omega/(2k/p_0)]^2}}{1-e^{-[k/2p_0-\beta\omega/(2k/p_0)]^2}}\right|.
\end{eqnarray}
For the real part $R(\omega,k)$ we use well-known spectral theorem
\begin{eqnarray}\label{R_omega}
R(\omega,k)=\textrm{v.p.} \int^{\infty}_{-\infty}\frac{d
\omega'}{\pi}\frac{I(\omega',k)}{\omega'-\omega}.
\end{eqnarray}
The spectrum of collective modes is determined by the equation
$1+\nu_kR(E_k,k)=0$. Using Eqs.~(\ref{I_omega}), (\ref{R_omega}) it is easy
to obtain the asymptotic formula for $R(\omega,k)$ \cite{Hore_Frankel} and to argue that
$E_k$ tends to the Bogoliubov spectrum when $an^{1/3}\gg1$.

\end{document}